\documentclass{moriond}
\usepackage{wrapfig}
\usepackage{caption}

\newcommand{\babar}{\mbox{\slshape B\kern-0.1em{\footnotesize A}\kern-0.1em B\kern-0.1em{\footnotesize A\kern-0.1em R}}}

\def\Journal#1#2#3#4{{#1} {\bf #2}, #3 (#4)}

\def\NCA{\em Nuovo Cimento}

\def\NIMA{{\em Nucl. Instrum. Methods} A}

\def\PRL{\em Phys. Rev. Lett.}
\def\PRD{{\em Phys. Rev.} D}

\def\be{\begin{equation}}
\def\ee{\end{equation}}
\def\bea{\begin{eqnarray}}
\def\eea{\end{eqnarray}}

\def\fb{fb$^{-1}$}

\newcommand{\Photo}{\includegraphics[height=35mm]{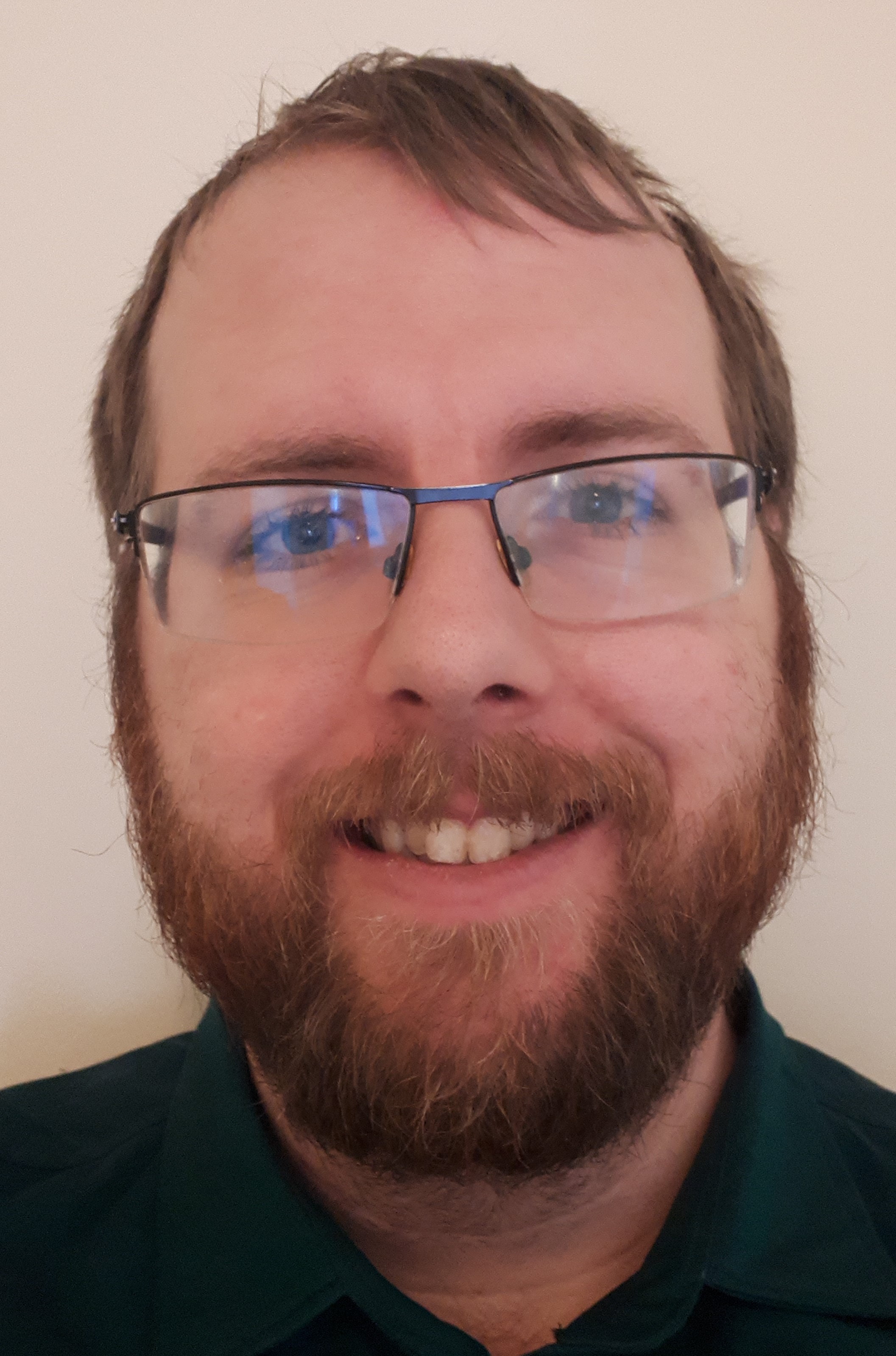}}

\begin{document}
\vspace*{4cm}
\title{Precision measurement of the ratio $\mathcal{B}$$\left(\Upsilon(3S)\rightarrow\tau^+\tau^-\right)/\mathcal{B}$$\left(\Upsilon(3S)\rightarrow\mu^+\mu^-\right)$}

\author{Caleb Miller \\ \small{on behalf of the \babar~Collaboration}} 

\address{University of Victoria, 3800 Finnerty Rd., Victoria V8P 5C2, Canada}

\maketitle\abstracts{
The \babar~collaboration has measured the ratio $R_{\tau\mu}^{\Upsilon(3S)}=\mathcal{B}(\Upsilon(3S)\rightarrow\tau^+\tau^-)/\mathcal{B}(\Upsilon(3S)\rightarrow\mu^+\mu^-)$ with a high level of precision. This measurement utilized a 28 fb$^{-1}$ dataset collected at a center-of-mass energy of 10.355 GeV. The measured ratio, $R_{\tau\mu}^{\Upsilon(3S)}$, is measured to be $R_{\tau\mu}^{\Upsilon(3S)}=0.966\pm0.008_{\textrm{stat}}\pm0.014_{\textrm{syst}}$. This value is within 2 standard deviations of the standard model prediction $R_{\tau\mu}^{\Upsilon(3S)}$=0.9948. The new measurement is approximately a factor of 6$\times$ more precise than the only prior measurement. This increased precision is in part due to a more complete analysis of the radiative tail in the $\Upsilon(3S)$ decay, in addition to a significant increase in statistics.
}

In the Standard Model the decay width of a bound state of a quark and an antiquark into a pair of leptons is expressed as\cite{Rfactor}:
\begin{equation}
	\Gamma=4\alpha^2Q^2\frac{|R(0)|^2}{M^2}\left(1+2\frac{m_{\ell}^2}{M^2}\right)\sqrt{1-4\frac{m_{\ell}^2}{M^2}},
\end{equation}
where $\Gamma$ is the decay width, $\alpha$ is the fine structure constant, $Q$ is the quark charge, $R(0)$ is the value of the radial wave function evaluated at the origin, $M$ is the resonance mass and $m_{\ell}$ is the lepton mass. By taking the ratio of the tau and muon decay widths, $R_{\tau\mu}$:
\begin{equation}
	R_{\tau\mu}=\frac{\Gamma_{\Upsilon\rightarrow\tau\tau}}{\Gamma_{\Upsilon\rightarrow\mu\mu}}=\frac{(1+2m^2_\tau/M^2)\sqrt{1-4m^2_\tau/M^2}}{(1+2m^2_\mu/M^2)\sqrt{1-4m^2_\mu/M^2}}.
\end{equation}
It can be seen that the ratio $R_{\tau\mu}$ is independent of the hadronic uncertainties and close to 1 due to the heavy mass of the $\Upsilon$ resonances. Leptonic decays of the $\Upsilon(nS)$ mesons are good candidates to search for phenomena beyond the SM. For example, measuring this ratio could shed light on the hint for new physics seen in $B\rightarrow D^{(*)}\tau\nu/ B\rightarrow D^{(*)}\ell\nu$ $(\ell=e,\mu)$\cite{SMR}. This ratio has measured once before by the CLEO collaboration, 	$R_{\tau\mu}=1.05\pm0.08\pm0.05$\cite{CLEO}. We present a novel technique which takes advantage of the differences between resonant and off-resonant di-muon processes to achieve a new level of precision. In the resonant process, $e^+e^-\rightarrow\Upsilon(3S)\rightarrow\mu^+\mu^-$, initial-state radiation(ISR) is highly suppressed compared to the background continuum process, $e^+e^-\rightarrow\mu^+\mu^-$. Our approach allows us to perform the measurement in a way that fully accounts for these radiative effects and does not require precise knowledge of the luminosity.\\
The data used in this analysis was collected by the \babar~detector at the asymmetric-energy $e^+e^-$ collider PEP-II. In particular we make use of 122 million $\Upsilon(3S)$ decays from an integrated luminosity of 27.96\fb. In addition to this analysis sample, three control samples are used: 78.3\fb~collected at the $\Upsilon(4S)$ resonance, 7.75\fb~collected 40 MeV below the $\Upsilon(4S)$ resonance, and 2.62\fb~collected 30 MeV below the $\Upsilon(3S)$ resonance. All of the data samples used were collected in 2007 and 2008 following a final upgrade in order to ensure we have a consistent detector configuration. The \babar~detector has been described in detail previously\cite{BBar,BBarUpgrade}. \\
Muon pairs are selected by requiring two and only two high momentum collinear (opening angle$> 160^\circ$) charged tracks with energy depositions consistent with a muon hypothesis. The scaled invariant mass, $M_{\mu\mu}/\sqrt{s}$, is required to be in the range $0.8<M_{\mu\mu}/\sqrt{s}<1.1$, which according to Monte-Carlo (MC) studies results in a 99.9\% purity. The tau pairs are selected from decays where both tau's decay to a single charged particle. One of the charged tracks is required to be an electron as identified by the \babar~particle identification, while the other track must fail the same electron definition. To further reduce backgrounds in the tau sample the tracks are required to be separated by at least $100^\circ$ in the centre-of-mass frame. The total measured energy must be less than $0.7\sqrt{s}$, and the acollinearity angle between the two tracks in the azimuthal plane must be greater than $3^\circ$. Furthermore we require the missing mass, $M_{miss}$, to satisfy $|M^2_{miss}/s|>0.01$, the missing momentum vector to point towards a sensitive region of the \babar~detector, and finally to reduce two-photon backgrounds the correlations in transverse momentum of the two charged tracks is exploited. This results in a $\tau^+\tau^-$ sample of 99\% purity.\\
\begin{wrapfigure}{R}{7cm}
	\centering
	\includegraphics[width=\linewidth]{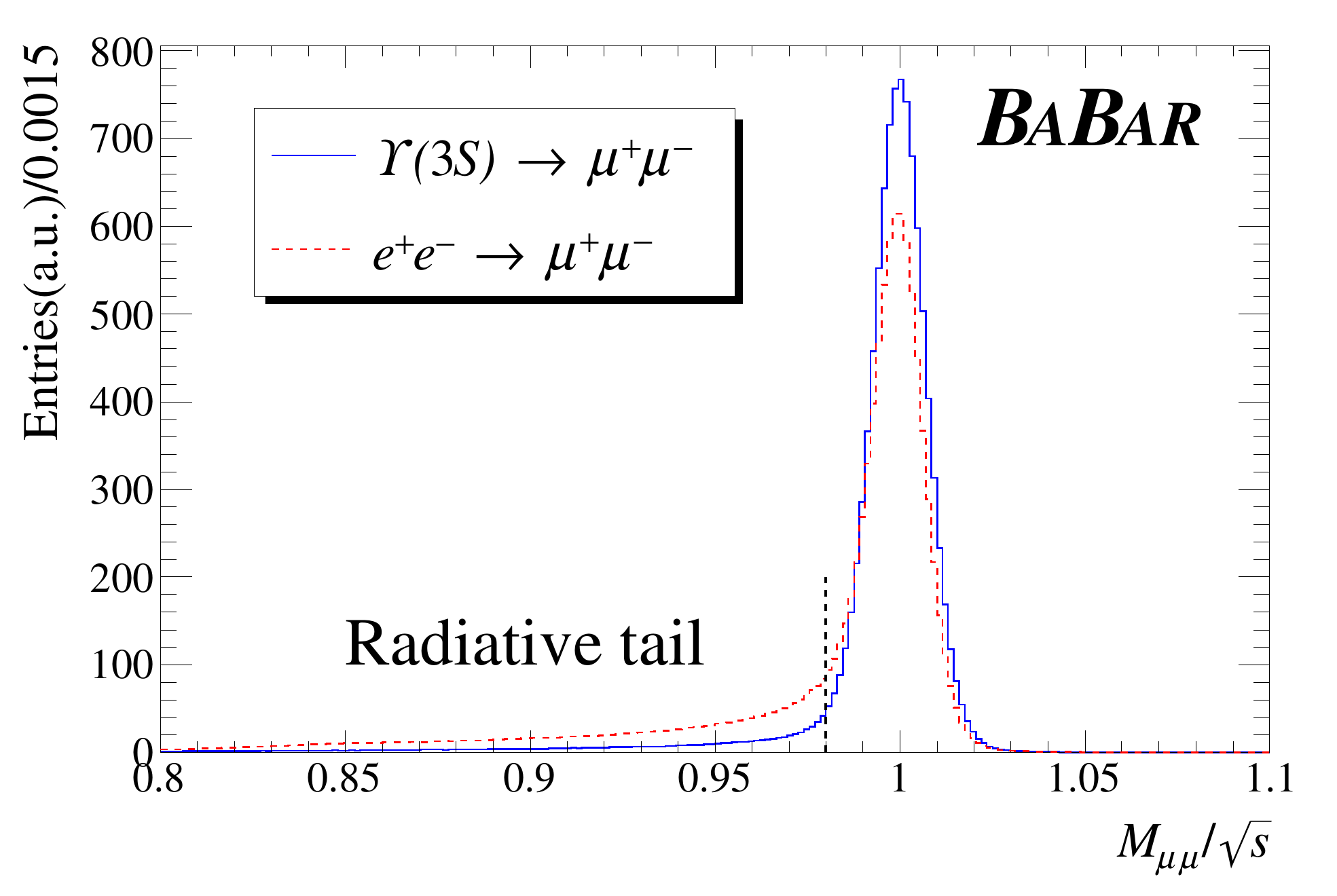}
	\caption{$M_{\mu\mu}/\sqrt{s}$ distribution for on-resonance $\Upsilon(3S)$ decays from MC and continuum di-muon production taken from off-resonance data. The distributions are normalized to the same number of events. The vertical dashed line defines the tail region to be $M_{\mu\mu}/\sqrt{s}<0.98$.}
	\label{fig:tail}
\end{wrapfigure}

The off-resonance data samples are used to correct for differences in the selection efficiency between data and MC. The $N_{\tau\tau}/N_{\mu\mu}$ ratios in the off-resonance $\Upsilon(3S)$ and $\Upsilon(4S)$ data are $0.11665\pm0.00029$ and $0.11647\pm0.00017$ respectively. The high level of agreement demonstrates that the efficiency is independent of centre-of-mass energy. The correction to the MC efficiency estimation is given by:
\begin{eqnarray}
		C_{MC}=(\varepsilon_{\tau\tau}/\varepsilon_{\mu\mu})^{\textrm{data}}/(\varepsilon_{\tau\tau}/\varepsilon_{\mu\mu})^{\textrm{MC}} \nonumber \\
		=1.0146\pm0.0016.
\end{eqnarray}
The amount of data collected off-resonance is an order of magnitude lower than the $\Upsilon(3S)$ data. In order to measure the continuum contribution in a way that is not statistically limited we also use on-resonance $\Upsilon(4S)$ data, a sample with an integrated luminosity of 78.3\fb. The leptonic decay width of the $\Upsilon(4S)$ is 1.57$\times10^{-5}$ of the total width, which results in a negligible amount of resonance-produced dilepton events remaining in the selection. \\
\begin{figure}[ht]
	\includegraphics[width=0.45\linewidth]{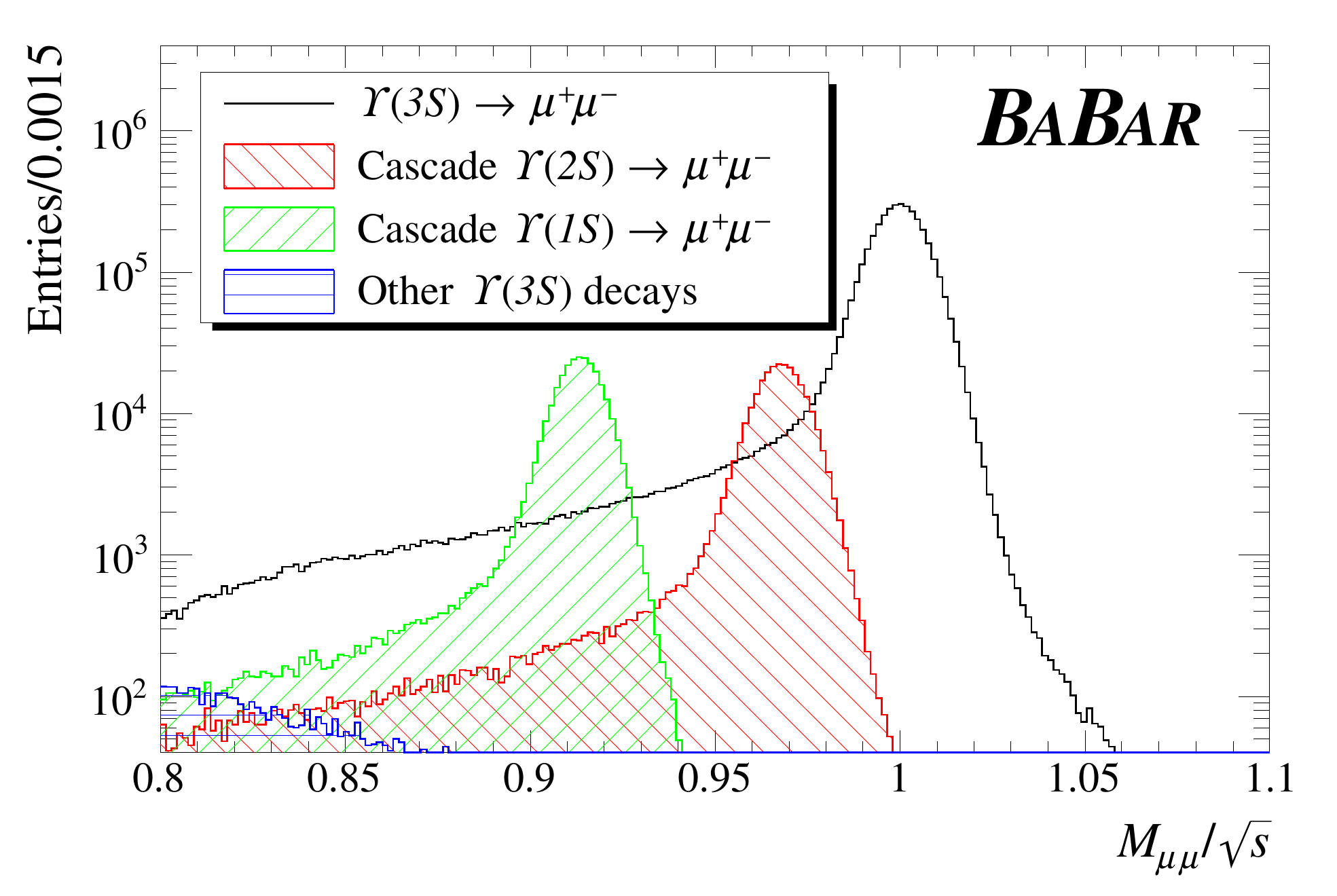}
	\includegraphics[width=0.45\linewidth]{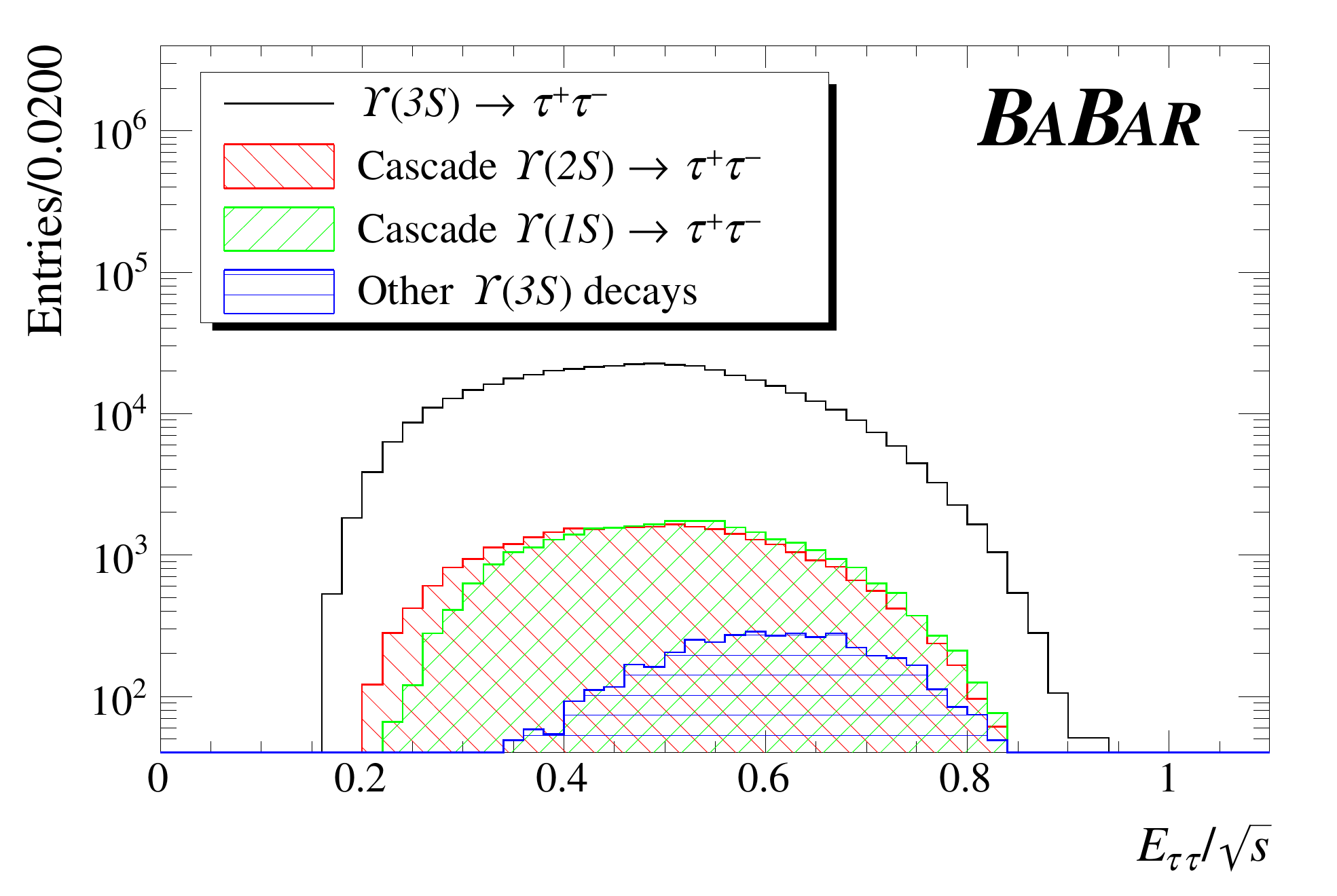}
	\caption{$M_{\mu\mu}/\sqrt{s}$(left) and $E_{\tau\tau}/\sqrt{s}$(right) distributions in MC. Note that in the di-muon distribution the cascade decays are clearly visible while being indistinguishable in the di-tau distribution.}
	\label{fig:shapes}
\end{figure}
\begin{wrapfigure}{R}{7cm}
	\includegraphics[width=\linewidth]{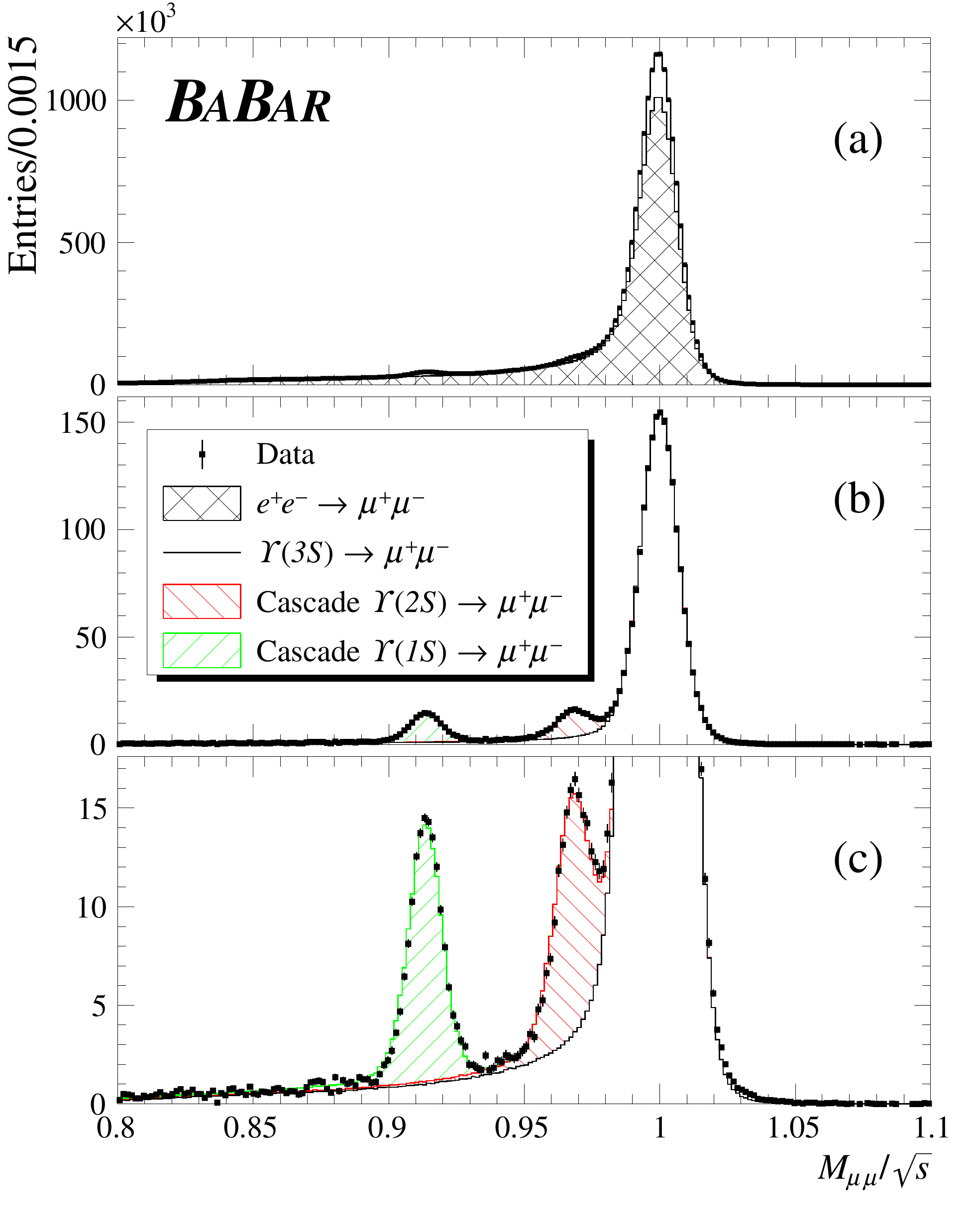}
	\caption{The results of the fit to the $\Upsilon(3S)$ in $M_{\mu\mu}/\sqrt{s}$. In (a) all events are shown, in (b) the continuum is hidden, and (c) is a magnified view of (b) on the cascade decays in the radiative tail.}
	\label{fig:mufit}
\end{wrapfigure}
We are able to discriminate between $\Upsilon(3S)\rightarrow\mu^+\mu^-$ and the continuum $e^+e^-\rightarrow\mu^+\mu^-$ by exploiting the effect ISR has on the two processes. If an ISR photon is more than a few MeV in energy the $e^+e^-$ interaction no longer has the energy to produce the $\Upsilon(3S)$ bound state. This results in on-resonance and continuum events having a significant difference in the relative number of events in the radiative tail of the invariant mass distribution. This can be seen in Fig. \ref{fig:tail} where 23\% of continuum events are in the radiative tail ($M_{\mu\mu}/\sqrt{s}<0.98$) compared to 7\% for resonance decays. This difference in the relative abundance of each region lends itself to a binned maximum-likelihood fit using a Barlow\&Beeston template fit methodology\cite{Barlow}. \\

This fitting method requires templates of the distribution of each source contributing to the fitted variable. The continuum template is taken from the data control samples as previously described, while the on-resonance $\Upsilon(3S)$ production is taken from MC. In addition to the $\Upsilon(3S)$ process we also account for the radiative return process where through ISR $\Upsilon(nS)\rightarrow\ell^+\ell^-$ `cascade' processes occur. These radiative return processes have been studied in detail by \babar\cite{radRet}. In addition to being used in the fit, we account for these processes to correct the continuum template taken from the $\Upsilon(4S)$ on-resonance data. We preform the fit simultaneously on both the $M_{\mu\mu}/\sqrt{s}$ and $E_{\tau\tau}/\sqrt{s}$ distributions. The shape of the $\Upsilon(3S)$ components are shown in Fig.~\ref{fig:shapes}. The fit is preformed with two free parameters, the number of $\Upsilon(3S)\rightarrow\mu^+\mu^-$ events ($N_{\mu\mu}$), and the ratio $\tilde{R}=N_{\tau\tau}/N_{\mu\mu}$. For the non-signal distributions, this ratio is fixed either to data in the case of continuum or MC for the cascade distributions. \\

\begin{figure}[h]
	\centering
	\begin{minipage}{0.45\textwidth}
		\centering
		\includegraphics[width=\linewidth]{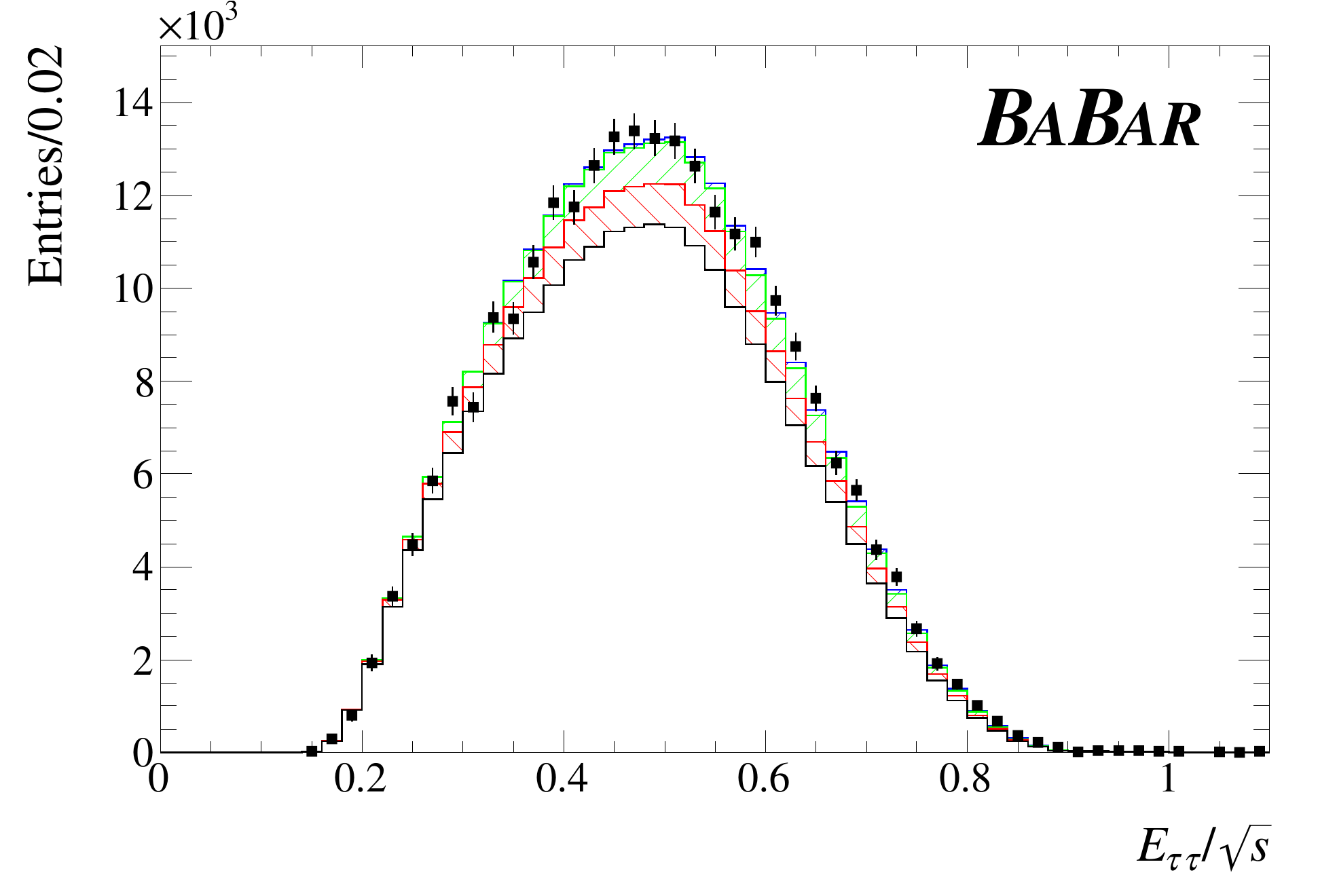}
		\caption{The results of the fit to the $\Upsilon(3S)$ in $E_{\tau\tau}/\sqrt{s}$ with the continuum fit subtracted. Legend is the same as in Fig.\ref{fig:mufit}.}%
		\label{fig:taufit}
	\end{minipage}
	\begin{minipage}{0.45\textwidth}
		\centering
		\captionsetup{type=table}
		\vspace{-15pt}
		\caption{The list of estimated systematic uncertainties}%
		\label{tab:sys}
		\begin{tabular}{|lc|}\hline
			Source & Uncertainty(\%) \\ \hline
			Particle identification & 0.9 \\
			Cascade decays & 0.6 \\
			Two-photon production & 0.5 \\
			$\Upsilon(3S)\rightarrow$hadrons & 0.4 \\
			MC shape & 0.4 \\
			$B\bar{B}$ & 0.2 \\
			ISR & 0.2 \\ \hline
			Total & 1.4 \\ \hline
		\end{tabular}
		
	\end{minipage}
\end{figure}

Figures \ref{fig:mufit} and \ref{fig:taufit} show the distributions with scaling for each contribution set by the fit. The fit gives $\tilde{R}=N_{\tau\tau}/N_{\mu\mu}=0.10788\pm0.00091$ and $N_{\mu\mu}=(2.014\pm0.015)\times10^6$. In order to construct the final ratio, $R_{\tau\mu}$, we account for two more contributions. First the relative selection efficiency of the muon and tau pairs. Our studies of the MC show $\varepsilon_{\tau\tau}/\varepsilon_{\mu\mu}=0.11041\pm0.00015$. The final factor to be considered is a small amount of $\Upsilon(4S)\rightarrow B\bar{B}$ which mimic tau-pair events. According to our MC studies the $B\bar{B}$ contribution to the muon distributions is negligible, however in the tau distribution a correction of $\delta_{B\bar{B}}=0.42\%$ is required. This gives us a final ratio of:
\begin{equation}
	R^{\Upsilon(3S)}_{\tau\mu}=\tilde{R}\frac{1}{C_{MC}}\frac{\varepsilon_{\mu\mu}}{\varepsilon_{\tau\tau}}(1+\delta_{B\bar{B}})=0.9662\pm0.0084.
\end{equation}
Dominant systematic uncertainties are listed in Table \ref{tab:sys}. These systematics are added in quadrature to give a total systematic uncertainty of 1.4\%

In conclusion, using data collected at and near the $\Upsilon(3S)$ and $\Upsilon(4S)$ resonances, \babar~has measured the ratio of the tau and muon leptonic branching fractions of the $\Upsilon(3S)$ to be:
\begin{equation}
	R^{\Upsilon(3S)}_{\tau\mu}=0.966\pm0.008_{\textrm{stat}}\pm0.014_{\textrm{syst}}.
\end{equation}
This result is six times more precise than the only prior measurement\cite{CLEO} and is within two standard deviations of the SM prediction of 0.9948\cite{SMR}. These results have been published in PRL\cite{Paper}, where additional details can be found.\\ 

\end{document}